\documentclass[aps,preprint,tightenlines]{revtex4}
\usepackage{amsfonts}
\usepackage{amsmath}
\usepackage{amssymb}
\usepackage{graphicx}

\input{tcilatex}

\begin{document}

\preprint{quant/ph}
\title[quantum vacuum spacecraft ]{A Gedanken spacecraft that operates using
the quantum vacuum (Dynamic Casimir effect) }
\author{G. Jordan Maclay}
\email{jordanmaclay@quantumfields.com}
\affiliation{Quantum Fields LLC, Richland Center WI 53581}
\author{Robert. L. Forward }
\affiliation{Forward Unlimited LLC, Clinton WA 98236}

\begin{abstract}
Conventional rockets are not a suitable technology for deep space missions.
\ Chemical rockets require a very large weight of propellant, travel very
slowly compared to light speed, and require significant energy to maintain
operation over periods of years. \ For example, the 722 kg Voyager
spacecraft required 13,600 kg of propellant to launch and would take about
80,000 years to reach the nearest star, Proxima Centauri, about 4.3 light
years away. \ There have been various attempts at developing ideas on which
one might base a spacecraft that would permit deep space travel, such as
spacewarps. \ \ In this paper we consider another suggestion from science
fiction and explore how the quantum vacuum might be utilized in the creation
of a novel spacecraft. \ The spacecraft is based on the dynamic Casimir
effect, in which electromagnetic radiation is emitted when an uncharged
mirror is properly accelerated in the vacuum. \ The radiative reaction
produces a dissipative force on the mirror that tends to resist the
acceleration of the mirror. \ This force can be used to accelerate a
spacecraft attached to the mirror. \ We also show that, in principal, one
could obtain the power to operate the accelerated mirror in such a
spacecraft using energy extracted from the quantum vacuum using the standard
Casimir effect with a parallel plate geometry. \ Unfortunately the method as
currently conceived generates a miniscule thrust, and is no more practical
than a spacewarp, yet it does provide an interesting demonstration of our
current understanding of the physics of the quantized electromagnetic field
in vacuum. \ 
\end{abstract}

\volumeyear{year}
\volumenumber{number}
\issuenumber{number}
\eid{identifier}
\date{February 22,2003}
\received[Received text]{date}
\revised[Revised text]{date}
\accepted[Accepted text]{date}
\published[Published text]{date}
\startpage{1}
\endpage{2}
\maketitle

\section{ \ INTRODUCTION}

Our objective in this paper is to explore how one might make use of the
properties of the quantum vacuum in the creation of a spacecraft. \ Rockets
employing chemical or ionic propellants require the transport of
prohibitively large quantities of propellant. \ If the properties of the
quantum vacuum could somehow be utilized in the production of thrust, that
would provide a decided advantage since the vacuum is everywhere. \ The
three milestones for deep space or interstellar travel in the NASA
Breakthrough Propulsion Physics (BPP) Project are: 1) the development of
propellantless propulsion; 2) the development of methods that reduce travel
time by orders of magnitude; and 3) new methods of providing the energy to
operate the spacecraft\cite{bpp}. \ At this embryonic stage in the
exceedingly brief history of interstellar spacecraft, we are attempting in
this paper to distinguish between what is possible and what is not possible
within the context of our current understanding of the relevant physics. \
Science fiction writers have written about the use of the quantum vacuum for
spacecraft for decades but no research has validated this suggestion. \
Arthur C. Clark, who proposed geosynchonous communications satellites in
1945, described a"quantum ramjet drive" in 1985 in "Songs of Distant Earth",
and observed in the Acknowledgement, "If vacuum fluctuations can be
harnessed for propulsion by anyone besides science-fiction writers, the
purely engineering problems of interstellar flight would be solved\cite%
{clarke}."

\ Since we do not know how to harness the energy of vacuum fluctuations, we
have focused on a slightly different approach in this paper, and have
considered how we might transfer momentum to the vacuum from the spacecraft.
\ The spacecraft proposed in this paper, which addresses the first and last
BPP milestones, is described as a "gedanken spacecraft" since its design is
not intended as an engineering guide, but just to illustrate possibilities.
\ Indeed based on our current understanding of quantum vacuum physics, one
could reasonably argue that a better gedanken spacecraft could be propelled
by oscillating a charged mirror or simply using a flashlight or laser to
generate photons. \ Although the performance of the gedanken spacecraft as
presented is disappointing and is no more practical than a spacewarp\cite%
{thorne}, it illustrates many current notions about the quanum vacuum, and
it is interesting to understand the potential role of quantum vacuum
phenomena in a macroscopic system like space travel. \ Obviously, we need
some major breakthroughs in our understanding if we are to realized the
dream of space travel as presented in science fiction works.

In 1948, when physicists were developing new ideas in quantum field theory,
Lamb made a seminal measurement showing a shift in the energy levels of the
hydrogen atom that was interpreted as due to the interaction of the atom
with the quantum fluctuations of the vacuum state of the\ quantized
electromagnetic field\cite{lamb}. \ Theory showed this field produced a
complex radiative shift that shifted both the energy level and its lifetime%
\cite{bethe}. \ Independently, Casimir made a prediction of an attractive
quantum vacuum force that would exist between two parallel, uncharged, metal
plates due to the boundary conditions the quantized fluctuating vacuum
electromagnetic field must meet on the surface of the plates\cite{cas1}. \
About twelve years later, Lifshitz provided another explanation of the
Casimir effect based on fluctuations in the very long range behavior of the
intermolecular potentials. \ Five decades passed before there was
experimental verification in the landmark experiments of Lamoroux \cite%
{lamoroux}using a torsion pendulum, and Mohideen\cite{mohideen} using an
Atomic Force Microscope (AFM). \ In the AFM measurements a 200 $\unit{%
\mu%
m}$ sphere, metallized with gold, is attached to the end of a cantilever,
and moved near a flat surface. \ The deflection of the cantilever is
measured optically, allowing the detection of n$\unit{N}$ forces. \ The AFM
measurements have become increasingly precise and corrections due to
conductivity, temperature, and surface roughness have been developed, giving
agreement between theory and experiment at about the 1 \% level\cite{klim}.
\ \ Current efforts are designed to explore the temperature dependence of
the Casimir force\cite{mohideenpc}. \ Proposals regarding the appearance of
and the use of Casimir forces in micromechanical devices are being realized
in microelectromechanical systems \cite{serry}\cite{serry2}. \ \ Buks et al
have measured adhesion energies due to Casimir forces using gold films\cite%
{buks}. \ \ Casimir forces have recently been used to actuate a
microelectromechanical torsion device\cite{chan}. \ The inverse fourth power
behavior of the parallel plate Casimir force has been exploited to make this
device operate as an anharmonic oscillator which serves as a sensitive
position sensor\cite{chan2}. \ The concept of a vacuum force that arises
from the application of boundary condition has been generalized to various
geometries and topologies, and has been reviewed\cite{pwm}\cite%
{plunienreview}\cite{bordag}.

The experimental verification of Casimir's prediction is often cited as
proof of the reality of the vacuum energy density of quantum field theory. \
Yet, as Casimir himself observed, other interpretations are possible:

\begin{quotation}
The action of this force [between parallel plates] has been shown by clever
experiments and I\ think we can claim the existence of the electromagnetic
zero-point energy without a doubt. But one can also take a more modest point
of view. \ Inside a metal there are forces of cohesion and if you take two
metal plates and press them together these forces of cohesion begin to act.
\ On the other hand you can start with one piece and split it. \ Then you
have first to break chemical bonds and next to overcome van der Waals forces
of classical type and if you separate the two pieces even further there
remains a curious little tail. The Casimir force, \textit{sit venia verbo},
is the last but also the most elegant trace of cohesion energy \cite{cas3}.
\end{quotation}

The nearly infinite zero-point field energy density is a simple and
inexorable consequence of quantum theory, but it brings puzzling
inconsistencies with another well verified theory, general relativity. \
Casimir effects have also been derived and interpreted in terms of \textit{%
source} fields arising from fluctuations within matter in both conventional 
\cite{pwm} and nonconventional \cite{schw} quantum electrodynamics. These
interpretations have spawned the evolution of two views which have proved to
be equivalent in the limited systems considered to date. \ In one view the
energy in the quantum vacuum is seen as a consequence of the fluctuations in
the electrical potential from nearby matter; \ in the other view the energy
in space is seen as due to the ground state of the quantized electromagnetic
field. \ No experiment yet proposed has been shown to clearly distinguish
between these two viewpoints, but to date only experiments on flat or nearly
flat surfaces have been done. \ Experiments with resonant cavities may
resolve some of these questions\cite{dodonov}. \ Theoretical calculations
are often done using the simplest approach or the one of personal
preference. \ In this paper, we will perform computations using the
quantized vacuum field model at zero Kelvin and refer to the properties of
the quantum vacuum, with the current understanding that an equivalvent
derivation could probably be based on effects which arise from the tails of
van der Waals forces between molecules. \ Intuitively one might conjecture
that space travel based on the properties of the quantum vacuum is more
likely to become a reality if there is actually fluctuating energy within
all space, and we are not just dealing with the tails of molecular
potentials of matter nearby.

Casimir effects result from changes in the ground-state fluctuations of a
quantized field that occur because of the particular boundary conditions. \
Casimir effects occur for all quantum fields and can also arise from the
choice of topology. In the special case of the vacuum electromagnetic field
with dielectric or conductive boundaries, various approaches suggest that
Casimir forces can be regarded as macroscopic manifestations of many-body
retarded van der Waals forces \cite{pwm}, \cite{Power}.

In 1970, Moore considered the effect of an uncharged one dimensional
boundary surface that moved, with the very interesting prediction that it
should be possible to generate real photons from this motion\cite{moore}. \
Energy conservation requires the existence of a radiation reaction force
working against the motion of the mirror\cite{netoandmachado}. \ The energy
expended moving the mirror against the radiative force goes into
electromagnetic radiation. \ This effect, generally referred to as the
dynamic or adiabatic Casimir effect, has been reviewed \cite{plunienreview}%
\cite{bordag}\cite{birrellanddavies}. \ The vacuum field exerts a force on
the moving mirror that tends to damp the motion. \ This dissipative force
may be understood as the mechanical effect of the emission of radiation
induced by the motion of the mirror. \ The Hamiltonian is quadratic in the
field operators, and formally analogous to the Hamiltonian describing photon
pair creation by parametric interaction of a classical pump wave of
frequency $\omega_{o}$ with a nonlinear medium\cite{jaekelandreynaud}. \
Pairs of photons with frequencies $\omega_{1}+\omega_{2}=\omega_{o}$ are
created out of the vacuum state. \ Furthermore the photons have the same
polarization, and the components of the corresponding wave vectors $%
\overrightarrow{k_{1}}$ and $\overrightarrow{k_{2}}$ taken along the mirror
surface must add to zero because of the translational symmetry:%
\begin{equation}
\overrightarrow{k_{1}}\cdot\widehat{x}+\overrightarrow{k_{2}}\cdot\widehat {x%
}=\omega_{1}\sin\theta_{1}+\omega_{2}\sin\theta_{2}=0
\end{equation}
\ \ This last equation relates the angles of emission of the photon pairs
with respect to the unit vector $\widehat{x},$ which is normal to the
surface. It is interesting that the photons emitted by the dynamic Casimir
effect are entangled photons. \ This analysis in terms of the effective
Hamiltonian is\ illuminating but not complete for perfect mirrors, because
no consistent effective Hamiltonian can be constructed in this case.

\ \ The dynamic Casimir effect was studied for\ a single, perfectly
reflecting mirror with arbitrary non-relativistic motion and a scalar field
in three dimensions by Ford and Vilenkin\cite{fordandvilenkin}. \ They
obtained expressions for the vacuum radiation pressure on the mirror. Barton
and Eberlein\ extended the analysis using a 1 dimensional scalar field to a
moving body with a finite refractive index\cite{bartonandeberlein}. \ The
vacuum radiation pressure and the radiated spectrum for a non-relativistic,
perfectly reflecting, infinite, plane mirror was computed by Neto and
Machado for the electromagnetic field in three dimensions, and shown to obey
the fluctuation-dissipation theorem from linear response theory\cite{neto}%
\cite{netoandmachado}. \ This theorem shows the fluctuations for a
stationary body yield information about the mean force experienced by the
body in nonuniform motion. Jaekel\ computed shifts in the mass of the mirror
for a scalar field in two dimension\cite{jaekel}. The mirror mass is not
constant, but rightfully a quantum variable because of the coupling of the
mirror to the fields by the radiation pressure. \ A detailed analysis was
done by Barton and Calogeracos for a dispersive mirror in 1 dimension, that
includes radiative shift in the mass of the mirror and the radiative
reaction force\cite{bartonandcal}. \ This model can be generalized to an
infinitesimally thin mirror\ with finite surface conductivity and a normally
incident electromagnetic field.

It might be of some interest to indicate why we analyze this spacecraft in
terms of the dynamic Casimir effect and not the Unruh Davies radiation. \
Historically, Unruh and Davies made the seminal investigations into the
thermal effects of acceleration on vacuum fluctuations\cite{unruh}\cite%
{davies}. \ They derived that a detector with constant acceleration $a$ for 
\textit{all} time (-infinity to +infinity) would observe black body
radiation at temperature $\hbar a/2\pi ck,$ where $k$ is Boltzmann's
constant. \ \ However, the renormalized vacuum expectation value of the
stress energy momentum tensor vanishes for the field with the zero velocity
detector, and by the transformation properties of tensors, it must also
vanish for the accelerated state. \ Hence the thermal quanta that excite the
accelerated detector have been described as 'fictitious' or 'quasi' thermal
quanta, although this conundrum may be more a reflection of the limited
meaning of the term 'particle' in accelerated reference frames or curved
spaces\cite{birrellanddavies}.\ \ It has been suggested that the thermal
spectrum appears to arise not from the "creation of particles" but the
distortion of the zero-point field\cite{hacyan}. \ Barton has observed that
unresolved difficulties still obstruct any clear understanding of accounts
of radiative processes referred to uniformly accelerated frames, and cites
the unphysical nature of uniform acceleration for all time\cite%
{bartonandeberlein}. \ Hence we have adopted the language and the methods of
analysis that characterize the dynamic Casimir effect in our analysis of the
gedanken spacecraft.

All calculations of electromagnetic fields and radiative forces to date on
moving plates have been for infinite plates. \ The contribution to the
radiative force on a vibrating infinite plate due to electromagnetic
radiation is the same for both sides of the plate\cite{neto}. \ From the
perspective of making a numerical estimate for a finite plate, this means
the theoretical results need to be adapted. \ \ To do this we assume that
the plate is large enough with respect to the wavelengths of the Fourier
components of the motion so that diffraction effects are small, and that the
force/area for the infinite plate is approximately the result for a finite
plate. \ Eberlein has made the very interesting suggestion that one could
calculate the force for a finite two sided plate as the zero eccentricity
limit of the force on an oblate spheroid\cite{claudiapc}\cite{eberleinson}.

Recently suggestions have been made\ to measure the motion induced radiation
escaping from a cavity in which one partially transmitting wall is vibrating%
\cite{reynaud}. The rate of photon emission is enhanced by the finesse of
the cavity and therefore may be orders of magnitude greater than for a
plate. \ Plunien and his collaborators showed that the rate of photon
emission by a vibrating cavity can be further enhanced by several orders of
magnitude if the temperature is several hundred degrees above zero Kelvin%
\cite{plunien}. \ The enhancement is proportional to the ground state\
photon population\ at the given temperature.

In Section \ref{sec2simmod}, we present a simplified analysis for the use of
the dynamic Casimir effect to accelerate a spacecraft. \ Also discussed is
the possibility of powering the accelerating mirror with energy extracted
from the quantum vacuum using the parallel plate Casimir force.\ In Section %
\ref{sec3detailed} a detailed description is given of a mirror trajectory
that produces a net impulse during each cycle of operation. \ After a
discussion of some possible methods to increase the thrust and impulse, the
conclusion follows.

\section{A SIMPLE MODEL FOR A QUANTUM VACUUM SPACECRAFT\label{sec2simmod}}

The important physical features of using the dynamic Casimir effect to
accelerate a spacecraft can be seen in a simplified, heuristic model. \
Assume that the spacecraft has an energy source, such as a battery, that
powers a motor that vibrates a mirror or a system of mirrors in a suitable
manner to generate radiation. \ We will assume that there are no internal
losses in the motor or energy source. \ We assume that at the initial time $%
t_{i}$, the mirrors are at rest. \ Then the mirrors are accelerated by the
motor in a suitable manner to generate a net radiative reaction on the
mirror, and at the final time $t_{f}$, the mirrors are no longer vibrating,
and the spacecraft has attained a non-zero momentum. \ \ We can apply the
first law of thermodynamics to the system of the energy source, motor, and
mirror at times $t_{i}$ and $t_{f}$ :%
\begin{equation}
\Delta Q=\Delta U+\Delta W
\end{equation}%
where $\Delta U$ represents the change in the internal energy in the energy
source, $-\Delta W$ represents the work done on the mirrors moving against
the vacuum, and $\Delta Q$ represents any heat transferred between the
system and the environment. \ We will assume that we have a thermally
isolated system and $\Delta Q=0$ so 
\begin{equation}
0=\Delta U+\Delta W
\end{equation}

By the conservation of energy, the energy $\Delta U$ extracted from the
battery goes into work done on the moving mirror $-\Delta W.$ \ Since the
mirror has zero vibrational kinetic energy and zero potential energy at the
beginning and the end of the acceleration period, and is assumed to operate
with no mechanical friction, all work done on the mirror goes into the
energy of the emitted radiation and the kinetic energy of the spacecraft of
mass $M:$%
\begin{equation*}
\Delta W=\Delta R+M\left( \Delta V\right) ^{2}/2
\end{equation*}
\ Thus the energy of the radiation emitted due to the dynamic Casimir effect
equals 
\begin{equation}
\Delta R=-\Delta U-M\left( \Delta V\right) ^{2}/2>0  \label{deltaR}
\end{equation}

The frequency of the emitted photons depends on the Fourier components of
the motion of the mirror. \ We assume that the radiant energy can be
expressed as a sum of energies of $n_{i}$ photons each with frequency $%
\omega _{i}$ :%
\begin{equation}
\Delta R=\dsum\limits_{i}n_{i}\hslash \omega _{i}
\end{equation}%
Let us assume that all photons are emitted normally from one side of the
accelerating surface. \ This assumption is not valid, as Eq.1 and the work
of Neto and Machado \cite{netoandmachado} clearly show, but it allows us to
obtain a best case scenario and illustrates the main physical features. \ If
all photons are emitted normally from one surface, then the total momentum
transfer $\Delta P$ \ is :%
\begin{equation}
\Delta P=\dsum\limits_{i}n_{i}\frac{\hslash \omega _{i}}{c}=\frac{\Delta R}{c%
}
\end{equation}%
where $c$ is the speed of light. \ Using Eq. \ref{deltaR}, we obtain the
result:%
\begin{equation}
\Delta P=\frac{-\Delta U}{c}-\frac{M\left( \Delta V\right) ^{2}}{2c}
\end{equation}

In a non-relativistic approximation $\Delta P=M\Delta V$ and the change in
velocity $\Delta V$ of the spacecraft is to second order in $\Delta U/Mc^{2}$%
:%
\begin{equation}
\frac{\Delta V}{c}=\frac{-\Delta U}{Mc^{2}}+\left( \frac{\Delta U}{Mc^{2}}%
\right) ^{2}
\end{equation}%
This represents a maximum change in velocity attainable by use of the
dynamic Casimir effect (or by the emission of electromagnetic radiation
generated by more conventional means) when the energy $\Delta U$ is
expended. \ The ratio $\Delta U/Mc^{2}$is expected to be a small number, and
we can neglect the second term in Eq. 8. (As a point of reference, \ for a
chemical fuel the ratio of the heat of formation to the mass energy is
approximately $10^{-10}$ .) \ With this approximation, we find the maximum
value of $\Delta V/c$ equals $\Delta U/Mc^{2},$ the energy obtained from the
energy source divided by the rest mass energy of the spacecraft. \ It
follows that the kinetic energy of the motion of the spacecraft $E_{ke}$ can
be expressed as%
\begin{equation}
E_{ke}=\frac{M(\Delta V)^{2}}{2}=\Delta U\frac{\Delta U}{2Mc^{2}}
\end{equation}%
This result shows that the conversion of potential energy $\Delta U$ from
the battery into kinetic energy of the spacecraft is an inefficient process
since\ $\Delta U/Mc^{2}$ is a small factor. \ Almost all of the energy $%
\Delta U$ has gone into photon energy. \ This inefficiency follows since the
ratio of momentum to energy for the photon is $1/c.$

\ In our derivation, no massive particles are ejected from the space craft
(propellantless propulsion ) and we have neglected: 1. the change in the
mass of the spacecraft as the stored energy is converted into radiation, 2.
radiative mass shifts, 3. complexities related to high energy vacuum
fluctuations and divergences, 4. all dissipative forces in the\ system\ used
to make the mirror vibrate. \ These assumptions are consistent with a
heuristic non-relativistic approximation. \ 

In this simplified model, we have not made any estimates about the rate of
photon emission and how long it would take to reach the maximum velocity.\ \
Rates are typically very low, typically $10^{-5}$ photons/sec \cite%
{netoandmachado}.

\subsection{Use of the static Casimir effect as an energy source \ \ }

In order to explore the possibility of a spacecraft that is based completely
on quantum vacuum properties, consider the use of an arrangement of
perfectly conducting, uncharged, parallel plates in vacuum as an energy
source. \ The Casimir energy $U_{C}(x)$ at zero degrees Kelvin between
plates of area $A$, separated by a distance $x$ is:%
\begin{equation}
U_{C}(x)=-\frac{\pi^{2}}{720}\frac{\hbar cA}{x^{3}}
\end{equation}

If we allow the plates to move from a large initial separation $a$ to a very
small final separation $b$ then the change in the vacuum energy energy
between the plates is approximately:%
\begin{align}
\Delta U_{C} & =U_{C}(b)-U_{C}(a) \\
& \approx-\frac{\pi^{2}}{720}\frac{\hbar cA}{b^{3}}
\end{align}

The attractive Casimir force has done work on the plates, and, in principal,
we can build a device to extract this energy with a suitable, reversible,
isothermal process, and use it to accelerate the mirrors. \ We neglect any
dissipative forces in this device, and assume all of the energy $\Delta
U_{C} $ can be utilized. \ Thus the maximum value of $\frac{\Delta V_{C}}{c}$
obtainable using the energy from the Casimir\ force "battery" is%
\begin{equation}
\frac{\Delta V_{C}}{c}=\frac{\pi^{2}}{720}\frac{1}{Mc^{2}}\frac{\hbar cA}{%
b^{3}}
\end{equation}

We can make an upper bound for this velocity by making further assumptions
about the composition of the plates. \ Assume that the plate of thickness $L$
is made of a material with a rectangular lattice that has a mean spacing of $%
d,$ and that the mass associated with each lattice site is $m$. \ Then the
mass of one plate is:%
\begin{equation}
M_{P}=AL\frac{m}{d^{3}}
\end{equation}

In principal, it is possible to make one of the plates in the battery the
same as the plate accelerated to produce radiation by the dynamic Casimir
effect. \ As the average distance between the plates is decreased, the
extracted energy is used to accelerate the plates over very small
amplitudes. \ If we assume we need to employ two plates in our spacecraft,
and that the assembly to vibrate the plates has negligible mass, then the
total mass of the spacecraft is $M=2M_{P}$ and we obtain an upper limit on
the increase in velocity:%
\begin{equation}
\frac{\Delta V_{C}}{c}=\frac{\pi^{2}}{1440}\frac{\hbar}{Lmc}\frac{d^{3}}{%
b^{3}}
\end{equation}

The final velocity is proportional to the Compton wavelength $(\hbar/mc)$ of
the lattice mass $m$ divided by the plate thickness $L$. \ Assume that the
final spacing between the plates is one lattice constant $(d=b),$ that the
lattice mass $m$ equals the mass of a proton $m_{p},$ and that the plate
thickness $L$ is one Bohr radius $a_{o}=\hbar^{2}/m_{e}e^{2}$, then we
obtain ($\alpha$ is the fine structure constant with approximate value of
1/137):%
\begin{equation}
\frac{\Delta V_{C}}{c}=\frac{\pi^{2}}{1440}\frac{\alpha m_{e}}{m_{p}}
\end{equation}

Substituting numerical values we find:%
\begin{equation}
\frac{\Delta V_{C}}{c}=\frac{\pi^{2}}{1440}\frac{1}{137}\frac{1}{1800}%
=2.\,\allowbreak78\times10^{-8}
\end{equation}

This corresponds to a disappointing final velocity of about $8$ $\unit{m}/%
\unit{s},$ about $10^{3}$ times smaller than for a large chemical rocket.\ \
As anticipated, the gedanken spacecraft is very slow despite the
unrealistically favorable assumptions made in the calculation, yet it does
demonstrate that it is possible to base the operation of a spacecraft
entirely on the properties of the quantum vacuum.

\subsection{Comments on the possible use of vacuum fluctuations as an energy
source}

Let us briefly comment on the long range possibility of extracting larger
amounts of energy from vacuum fluctuations. \ If, for example, it were
possible to make a structure equivalent to a parallel plate structure that
could have a final state with a separation 100 or 1000 times smaller than as
assumed above, then we would have the suggestion of a spacecraft operating
in the relativistic regime. \ Without a new approach, such a development is
probably not possible since the Casimir force decreases for separations less
than the plasma wavelength. \ (Several new approaches are discussed in
Section \ref{sectioninaccel}.) \ The reflectivity of a metal plate depends
on the applied frequency, and for frequencies above the plasma frequency the
reflectivity drops, causing the Casimir force\ to decrease. Alternatively,
if there were a device that could continuously extract energy from the
quantum fluctuations of the vacuum field, then this vacuum powered
spacecraft, and a lot of other unusual things, might become practical.

\ \ It is important to note that utilizing energy of the quantum
fluctuations of the electromagnetic field does not appear to directly
violate known laws of physics, based on the work of Forward and Cole and
Puthoff, however improbable or impossible such a development might seem \cite%
{forward}\cite{coleandputhoff}. \ Forward showed that it is possible to
conceive of a device, a foliated capacitor, in which one could extract
electrical energy from the quantum vacuum to do work. The energy is
extracted as the portions of the capacitor that repel each other due to
electrostatic forces come together under the influence of the Casimir force%
\cite{forward}. Cole and Puthoff used stochastic electrodynamics to examine
the process of removing energy from the vacuum fluctuations at zero
temperature from the viewpoint of thermodynamics and showed there is no
violation of the laws of thermodynamics\cite{coleandputhoff}. \ \ In the
same spirit, Rueda has suggested that very high energy particles observed in
space may derive their kinetic energy from a long term acceleration due to
the stochastic vacuum field\cite{rueda}. \ In a careful analysis, Cole has
shown that this process of energy transfer from the vacuum field to kinetic
energy of the particles does not violate the laws of thermodynamics\cite%
{colethermod}. \ In stochastic electrodynamics one treats the vacuum
fluctuations as a universal random classical electromagnetic field. \ A
formal analogy exists between stochastic electrodynamics and quantum
electrodynamics: \ the field correlation functions in one theory are related
to the Wightman functions in the other theory\cite{boyer}. \ Pinto has done
a calculation of a solid state Casimir device which is described as a
transducer of vacuum energy which can operate in a repetitive cycle\ \cite%
{pinto}. \ 

It is our opinion that further research is needed to gain a greater
understanding of what is possible and what are the fundamental limitations
and restrictions in processes involving energy transfer with the quantum
vacuum. \ One can take an alternative viewpoint here, that we are just
extracting energy from the tails of the molecular potentials, and that much
more energy would be available if we utilized the entire potential, for
example, by burning the material. \ \ However, this approach just leads us
back to the known limitations of chemical fuels. \ 

\section{DETAILED MODEL FOR PROPULSION USING DYNAMIC CASIMIR EFFECT\label%
{sec3detailed}}

\qquad Assume we have a flat, perfectly reflecting, mirror whose equilibrium
position is $x=0$. \ At a time $t$ where $t_{i}<t<t_{f}$ the location of the
mirror is given by $x(t).$ \ Neto has given an expression for the force per
unit area $F(t)$ on such a mirror\cite{neto}:%
\begin{equation}
F(t)=\lim_{\delta x\rightarrow0}\frac{\hbar c}{30\pi^{2}}\left[ \frac {1}{%
\delta x}\frac{d^{4}x(t)}{c^{4}dt^{4}}-\frac{d^{5}x(t)}{c^{5}dt^{5}}\right]
\label{F}
\end{equation}
where $\delta x$ represents the distance above the mirror at which the
stress-energy tensor is evaluated. \ The second term represents the
dissipative force that is related to the creation of travelling wave
photons, in agreement with its interpretation as a radiative reaction. \ In
computing the force due to the radiation from the mirror's motion, the
effect of the radiative reaction on $x(t)$ is neglected in the
nonrelativistic approximation. \ The divergent first term can be understood
in several ways. \ Physically it is a dispersive force that arises from the
scattering of low frequency evanescent waves. The divergence can be related
to the unphysical nature of the perfect conductor boundary conditions.
Forcing the field to vanish on the surface requires its conjugate momentum
to be unbounded. \ Thus the average of the stress-energy tensor $\langle
T_{\mu\nu}\rangle$ is singular at the surface for the same reason that
single-particle quantum mechanics would require a position eigenstate to
have infinite energy\cite{fordandvilenkin}. \ \ This divergent term can be
lumped into a mass renormalization, and therefore disappears from the
dynamical equations when they are expressed in terms of the observed mass of
the body\cite{bartonandeberlein}\cite{jaekel}. \ We will not discuss this
term further. \ We will assume that diffractions effects are small for our
finite plates.

The total energy radiated per unit plate area $E$ can be expressed as 
\begin{equation}
E=-\int\limits_{t_{1}}^{t_{2}}dtF(t)\frac{dx(t)}{dt}  \label{energy}
\end{equation}
Substituting Eq. \ref{F} for $F(t)$ we find%
\begin{equation}
E=\frac{\hbar}{30\pi^{2}c^{4}}\int\limits_{t_{1}}^{t_{2}}dt\left( \frac {%
d^{3}x(t)}{dt^{3}}\right) ^{2}  \label{E}
\end{equation}
The total impulse $I$ per unit plate area can also be computed as the
integral of the force\ per unit area over time:%
\begin{equation}
I=\int\limits_{t_{1}}^{t_{2}}dtF(t)=-\frac{\hbar}{30\pi^{2}}\frac{1}{c^{4}}%
\left( \frac{d^{4}x(t)}{dt^{4}}|_{t_{2}}-\frac{d^{4}x(t)}{dt^{4}}%
|_{t_{1}}\right)  \label{I}
\end{equation}
The total impulse $I$ equals the mass of the system $M$ per unit area times
the change in velocity $\Delta V$ in a non-relativistic approximation$:$ 
\begin{equation}
I=M\Delta V
\end{equation}

We want to specify a trajectory for the mirror that will give a net impulse.
\ One of the trajectories that has been analyzed is that of the harmonic
oscillator\cite{reynaud}\cite{fordandvilenkin}. \ In this case, the mirror
motion is in a cycle and we can compute the energy radiated per cycle per
unit area and the impulse per cycle per unit area. \ For a\ harmonic
oscillator of frequency $\Omega $ and period $T=2\pi /\Omega $, there is
only one Fourier component of the motion, so the total energy of each pair
of photons emitted is $\hbar \Omega =\hbar (\omega _{1}+\omega _{2}).$ \ For
a harmonically oscillating mirror the displacement is 
\begin{equation}
x_{ho}(t)=X_{o}\sin \Omega t
\end{equation}%
A computation based on Eqs. \ref{E} and \ref{I}\ shows there will be a net
power radiated in a cycle, however, the dissipative force for the harmonic
oscillator $F_{ho}$ will average to zero over the entire cycle as shown in
Fig. 1, so there will be no net impulse.\FRAME{ftbpFU}{2.3099in}{1.4209in}{%
0pt}{\Qcb{The displacement $x_{ho}$ and the radiative reaction force $F_{ho}$%
, bold line, for a harmonically oscillating mirror plotted as a function of
the time. For convenience $F_{ho}$ and $x_{ho}$ are normalized to unity
amplitude.}}{\Qlb{fig1}}{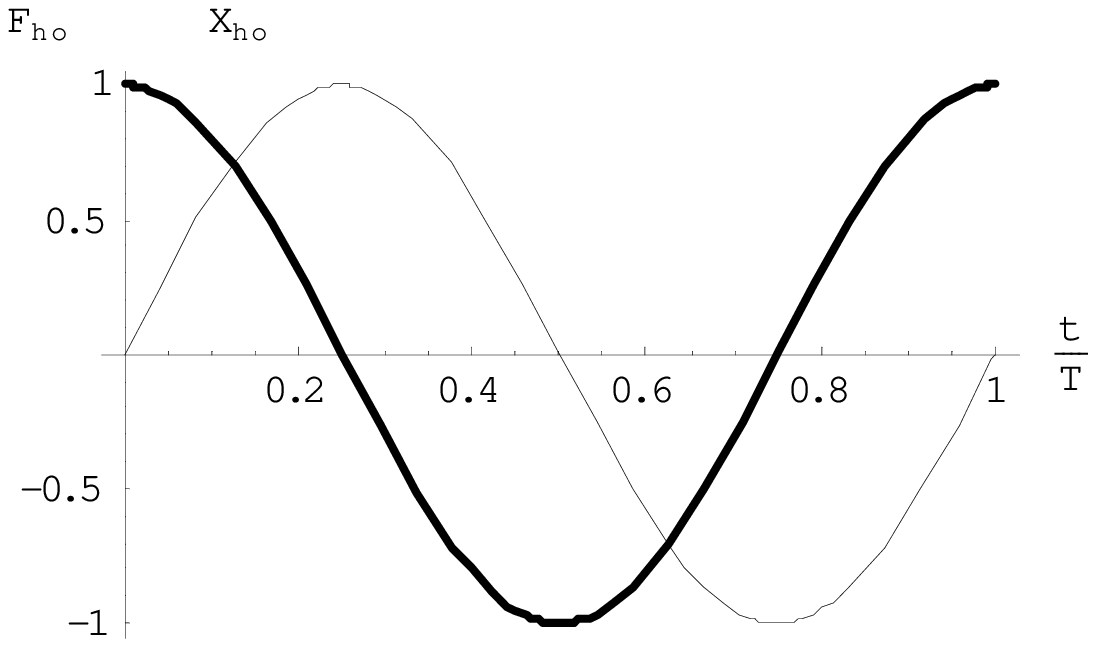}{\special{language "Scientific
Word";type "GRAPHIC";maintain-aspect-ratio TRUE;display "USEDEF";valid_file
"F";width 2.3099in;height 1.4209in;depth 0pt;original-width
4.5394in;original-height 2.7821in;cropleft "0";croptop "1";cropright
"1";cropbottom "0";filename 'maclayfig1.eps';file-properties "XNPEU";}}

\ In order to secure a net impulse, we need a modified mirror cycle. \ One
such model cycle can be readily constructed by using the harmonic function $%
x_{ho}(t)$ over the first and last quadrants of the cycle, where the force $%
F_{ho}$ is positive, and a cubic function $x_{c}(t)$ over the middle two
quadrants where $F_{ho}$ is negative:%
\begin{equation}
x_{c}(t)=\frac{X_{o}}{2}\frac{(\Omega t-\pi )^{3}}{(\pi /2)^{3}}-\frac{3X_{o}%
}{2}\frac{(\Omega t-\pi )}{\pi /2}
\end{equation}%
The coefficients for the cubic polynomial are chosen so that at $\Omega t=$ $%
\pi /2,3\pi /2$ the displacement and the first derivatives of $x_{c}(t)$ and 
$x_{ho}(t)$ are equal. As can be seen from Fig. \ref{fig2},\FRAME{ftbpFU}{%
2.3722in}{1.4589in}{0pt}{\Qcb{The normalized displacement for a harmonically
oscillating mirror $x_{ho}(t),$ solid line, and the cubic function $x_{c}(t)$%
, shown by the bold dots.}}{\Qlb{fig2}}{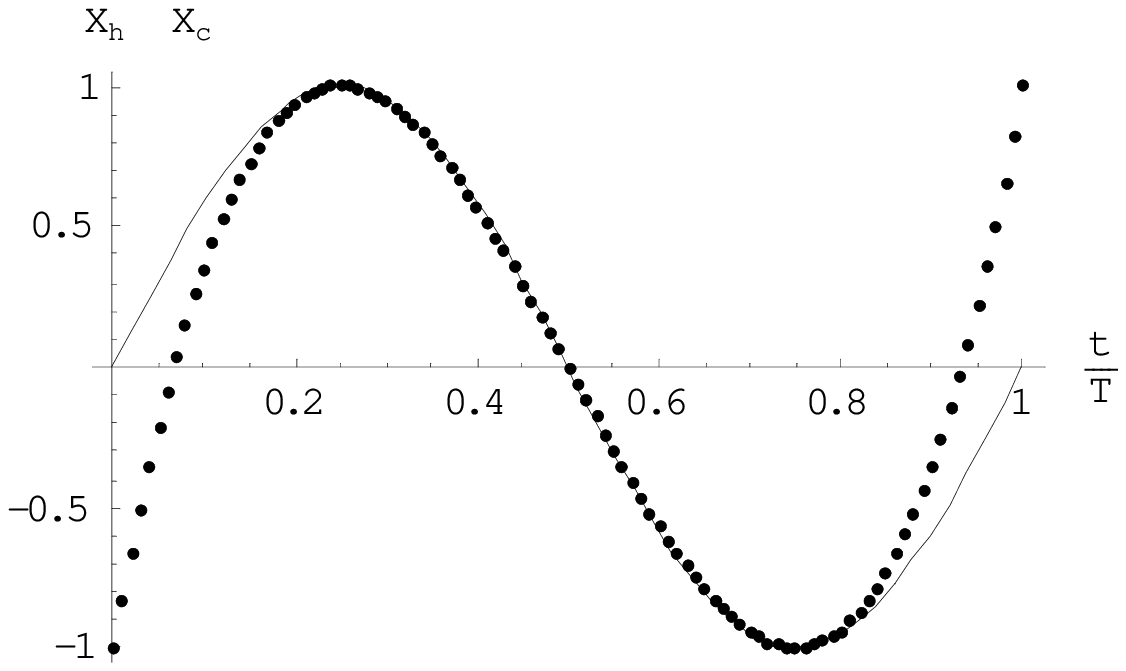}{\special{language
"Scientific Word";type "GRAPHIC";maintain-aspect-ratio TRUE;display
"USEDEF";valid_file "F";width 2.3722in;height 1.4589in;depth
0pt;original-width 4.5394in;original-height 2.7821in;cropleft "0";croptop
"1";cropright "1";cropbottom "0";filename 'maclayfig2.eps';file-properties
"XNPEU";}}the cubic function $x_{c}(t)$ matches \ $x_{ho}(t)$ quite closely
for $t$ in the interval $.25<t/T<.75.$ \ Of course the higher order
derivatives do not match, and that is precisely why the force differs. \ The
similarity in displacement and the difference in the resulting force is
striking. \ For the mirror displacement $x_{m}(t)$ in our model we choose: \
\ \ \ 
\begin{align}
x_{m}(t)& =x_{ho}(t)\text{ \ \ \ \ \ \ \ \ \ \ \ for }0\leq t/T\leq
0.25;0.75\leq t/T\leq 1 \\
x_{m}(t)& =x_{c}(t)\text{ \ \ \ \ \ \ \ \ \ \ \ \ \ for }0.25<t/T<0.75
\end{align}%
\ \FRAME{ftbpFU}{2.4241in}{1.4909in}{0pt}{\Qcb{\ The normalized displacement 
$x_{m}(t)$ and the corresponding normalized radiative force $F_{m}(t),$ \
bold solid line, are shown as functions of the time. \ The force is positive
in the first and last quarters, and zero in the middle half of the cycle.}}{%
\Qlb{fig3}}{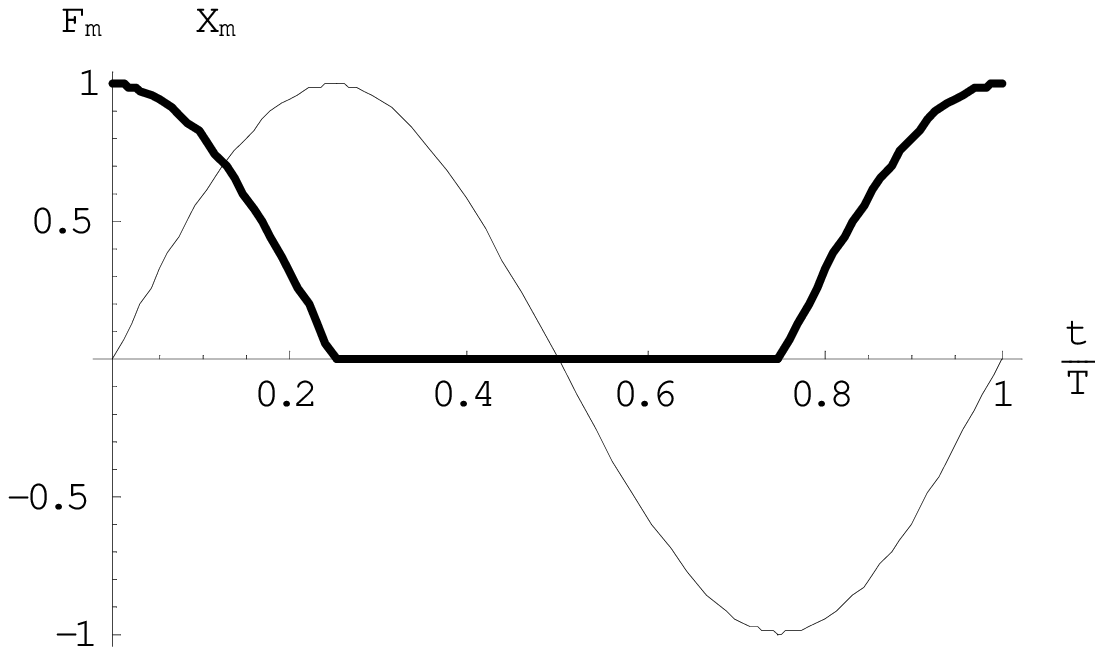}{\special{language "Scientific Word";type
"GRAPHIC";maintain-aspect-ratio TRUE;display "USEDEF";valid_file "F";width
2.4241in;height 1.4909in;depth 0pt;original-width 4.5394in;original-height
2.7821in;cropleft "0";croptop "1";cropright "1";cropbottom "0";filename
'maclayfig3.eps';file-properties "XNPEU";}}Fig. \ref{fig3} shows $x_{m}(t)$
plotted with the corresponding force per unit area $F_{m}(t)$ obtained from
Eq. \ref{F}. \ The force \ $F_{m}(t)$ is positive in the first and last
quarter of the cycle, and vanishes in the middle, where the trajectory is
described by the cubic. \ The energy radiated per area per cycle for our
model trajectory can be obtained from Eq. \ref{E}: 
\begin{equation}
E_{m}=-\frac{\hbar c}{60\pi }X_{o}^{2}(\frac{\Omega }{c})^{5}
\end{equation}%
The total impulse per area per cycle for our model $I_{m}$ trajectory is 
\begin{equation}
I_{m}=-\frac{\hbar }{15\pi ^{2}}X_{o}(\frac{\Omega }{c})^{4}
\end{equation}%
The impulse is first order in $\hbar ,$ and is therefore typically a small
quantum effect.\ \ Thus for our model cycle, the change in velocity per
second is $\Delta V_{m}/dt:$ \ 
\begin{equation}
\Delta V_{m}/dt=\frac{I_{m}\Omega }{M}
\end{equation}%
where $M$ is the mass per unit plate area of the spacecraft, and we assume
the plate is the only significant mass in the gedanken spacecraft. \ In
order to estimate$\ \Delta V_{m}$, we can make some further assumptions
regarding the mass of the plate per unit area. \ As before, we can make a
very favorable assumption regarding the mass per unit area of the plates $%
M=m_{p}/a_{o}^{2},$ which yields the change in velocity per second:%
\begin{equation}
\Delta V_{m}/dt=-\frac{\hbar }{15\pi ^{2}}X_{o}(\frac{\Omega }{c})^{4}\Omega 
\frac{a_{o}^{2}}{m_{p}}  \label{eqdvm}
\end{equation}%
If we substitute reasonable numerical values\cite{dodonov}\cite{reynaud}, a
frequency of $\Omega =3x10^{10}\unit{s}^{-1}\ $and an oscillation amplitude
of $X_{o}=10^{-9}\unit{m},\ $we find that $\Delta V_{m}/dt$ is approximately 
$3x10^{-20}\unit{m}/\unit{s}^{2}$ per unit area, not a very impressive
acceleration. \ \ Physically, one would imagine the surface of the mirror
vibrating with an amplitude of one nanometer. \ The limitation in the
amplitude arises because the maximum velocity of the boundary is
proportional to the elastic deformation, which cannot exceed about $10^{-2}$
for typical materials. \ The energy radiated per area $E_{m}\Omega $ is
about $10^{-25}\unit{W}/\unit{m}^{2}$ \ There are a number of methods to
increase these value by many orders of magnitude, as discussed below.

The efficiency of the conversion of energy expended per cycle in our model $%
E_{m}$ into kinetic energy of the spacecraft $E_{ke}=\frac{1}{2}M(\Delta
V_{m})^{2}$ $=I_{m}^{2}/2M$ is given in the nonrelativistic approximation by
the ratio:%
\begin{equation}
\frac{E_{ke}}{E_{m}}=\frac{\hbar}{Mc}\frac{1}{\pi}(\frac{\Omega}{c})^{3}
\end{equation}
\ With our assumptions, the approximate value of this ratio is $10^{-26}$,
making this conversion an incredibly inefficient process.\ \ 

\subsection{Methods to Increase Acceleration\label{sectioninaccel}}

The dynamic Casimir effect has yet to be verified experimentally. \ Hence
there have been a number of interesting proposals describing methods
designed to maximize the effect so it can be measured. \ In 1994, Law
predicted a resonant response of the vacuum to an oscillating mirror in a
one-dimensional cavity\cite{law}. \ The behavior of cavities formed from two
parallel mirrors that can move relative to each other is qualitatively
different from that of single plates. \ For example it is possible to create
particles in a cavity with plates separating at constant velocity\cite%
{carlos}. \ The very interesting proposal by Lambrecht et al concludes that
if the mechanical oscillation frequency is equal to an odd integer multiple
of the fundamental optical resonance frequency, then the rate of photon
emission from a vibrating cavity formed with walls that are partially
transmitting with reflectivity $r_{1}$ and $r_{2}$, is enhanced by a factor
equal to the finesse $f=1/\ln (1/r)$ of the cavity\cite{reynaud}\cite%
{dodonov}. \ For semiconducting cavities with frequencies in the Ghz range,
the finesse, which equals can be $10^{9}$, giving our gedanken spacecraft an
acceleration of $3x10^{-11}\unit{m}/\unit{s}^{2}$ based on Eq. \ref{eqdvm}.
\ \ Plunien et al have shown that the resonant photon emission from a
vibrating cavity is further increased in the temperature is raised\cite%
{plunien}.\ \ \ Assume that one has a gedanken spacecraft with a vibrating
cavity operating at a elevated temperature providing a $10^{10}$ total
increase in the emission rate. \ This would result in an acceleration per
unit area of the plates of $3x10^{-10}\unit{m}/\unit{s}^{2},$ a radiated
power of about $10^{-15}\unit{W}/\unit{m}^{2},$and an efficiency $%
E_{ke}/E_{m}$ of about $10^{-16}.$ \ After 10 years of operation, the
gedanken spacecraft velocity would be approximately\ $0.1\unit{m}/\unit{s}$
, which is about 5 orders of magnitude less than the current speed of
Voyager, $17\unit{km}/\unit{s}$, obtained after a gravity assist maneuver
around Jupiter to increase the velocity.\ \ The burn-out velocity for
Voyager at launch in 1977 was $7.1\unit{km}/\unit{s}$\cite{boston}. \ The
numerical results for the model obviously depend very strongly on the
assumptions made. \ For example, if a material could sustain an elastic
strain greater than about $10^{-2}$ then a larger amplitude would be
possible. \ Perhaps the use of nanomaterials, such as carbon nanotubes would
allow a much larger effective deformation. \ If the amplitude was 1mm
instead of 1 nm, the gedanken spacecraft would warrant practical
consideration.

Eberlein has shown that density fluctuations in a dielectric medium would
also result in the emission of photons by the dynamic Casimir effect\cite%
{eberleinden}. \ This approach may ultimately be more practical with large
area dielectric surfaces driven electrically at high frequencies. \ More
theoretical development is needed to evaluate the utility of this method. \
Other solid state approaches may also be of value with further technological
developments. \ For example, one can envision making sheets of charge that
are accelerated in MOS type structures. \ Yablonovitch has pointed out that
the zero-point electromagnetic field transmitted through a window whose
index of refraction is falling with time shows the same phase shift as if it
were reflected from an accelerating mirror\cite{yablonovitch}. \ He
suggested utilizing the sudden change in refractive index that occurs when a
gas is photoionized or the sudden creation of electron-hole pairs in a
semiconductor, which can reduce the index of refraction from $\sim3.5$ to $0$
in a very short time. \ Using subpicosecond optical pulses, the phase
modulation can suddenly sweep up low-frequency waves by many octaves. \ By
lateral synchronization, the moving plasma front can act as a moving mirror
exceeding the speed of light. \ Therefore one can regard such a gas or
semiconductor slab as an observational window on accelerating fields, with
accelerations as high as $\sim10^{20}\unit{m}/\sec$\cite{yablonovitch}$.$ \
Accelerations of this magnitude will have very high frequency Fourier
components. \ Eq. \ref{eqdvm} shows that the impulse goes as the fourth
power and the efficiency as the third power of the Fourier component for an
harmonic oscillator which suggests that with superhigh accelerations, and
the optimum time dependence of the field, and optimum shape of the
wavefronts, one might be able to secure much higher fluxes of photons and a
much higher impulse/second, with a higher conversion efficiency. \ A
calculation by Lozovik et al suggests that by itself the accelerated plasma
method is not as effective as a resonant cavity in producing photons.\cite%
{lozovik} \ 

Ford and Svaiter have shown that it may be possible to focus the fluctuating
vacuum electromagnetic field\cite{Ford}. This capability might be utilized
to create regions of higher energy density. \ \ This might be of use in a
cavity in order to increase the flux of radiated photons. \ There may be
also enhancements due to nature of the index of refraction for real
materials. \ \ For example, \ Ford has computed the force between a
dielectric sphere, whose dielectric function is described by the Drude
model, and a perfectly reflecting wall, with the conclusion that certain
large components of the Casimir force no longer cancel. \ He predicts a
dominant oscillatory contribution to the force, in effect developing a model
for the amplification of vacuum fluctuations. \ Barton and Eberlein have
shown that for materials with a fixed index of refraction, the force for a
one dimensional scalar field goes as $\left[ (n-1)/n\right] ^{2}$, which
suggest the possibility that one might be able to enhance the force by
selection of a material with a small index\cite{bartonandeberlein}. \ \ An
improbable approach would be based on reducing the spacecraft mass by tuning
the radiative mass shift\cite{reynaud}\cite{bartonandcal}.\ 

\section{CONCLUSION}

One of the objectives in this paper is to illustrate some of the unique
properties of the quantum vacuum and how they might be utilized in the
practical application of a gedanken spacecraft. \ \ \ We have demonstrated
that it is possible in principal to cause a spacecraft to accelerate due to
the dissipative force an accelerated mirror experiences when photons are
generated from the quantum vacuum. \ Further we have shown that one could in
principal utilize energy from the vacuum fluctuations to operate such a
vibrating mirror assembly. \ The application of the dynamic Casimir effect
and the static Casimir effect may be regarded as a proof of principal, with
the hope that the proven feasibility will stimulate more practical
approaches exploiting known or as yet unknown features of the quantum
vacuum. \ In any event, the physics of quantum fluctuations and its
application has been explored. \ A model gedanken spacecraft with a single
vibrating mirror was proposed which showed a very unimpressive acceleration
due to the dynamic Casimir effect of about $3x10^{-20}\unit{m}/\unit{s}^{2}$
with a very inefficient conversion of total energy expended into spacecraft
kinetic energy. \ Employing a set of vibrating mirrors to form a parallel
plate cavity increases the output by a factor of the finesse of the cavity
yielding an acceleration of about $3x10^{-10}\unit{m}/\unit{s}^{2}$ and a
conversion efficiency\ of about $10^{-16}.$ \ After 10 years at this
acceleration, the spacecraft would be traveling at $0.1\unit{m}/\unit{s}.$ \
\ Although these results are very unimpressive, it is important to not take
our conclusions regarding the final velocity in our simplified models too
seriously. \ The choice of numerical parameters is a best guess based on
current knowledge and can easily affect the final result by 5 orders of
magnitude. \ In \ about 1900 an article was published in Scientific American
proving that it was impossible to send a rocket, using a conventional
propellant, to the moon. \ The result was based on the seemingly innocuous
assumption of a single stage rocket.

From the status of current research in Casimir forces, it is clear that we
understand little of the properties of the quantum vacuum for real systems
with real material properties. \ For example, there is no general agreement
with regard to the calculations of static vacuum forces for geometries other
than infinite parallel plates of ideal or real metals at a temperature of
absolute zero. \ For non-zero temperatures corrections for flat, real metals
are uncertain\cite{bezerra}\cite{genet}. \ There are fundamental
disagreements about the computation of vacuum forces for spheres or
rectangular cavities, about how to handle real material properties and
curvature in these and other geometries\cite{bartonspheresetc}. \ Indeed, it
is very difficult to calculate Casimir forces for these simple geometries
geometries and to relate the calculations to an experiment. \ Calculations
have yet to be done for more complex, and potentially interesting
geometries. \ The usual problems in QED, such as divergences due to
unrealistic boundary conditions, to curvature, to interfaces with different
dielectric coefficients, etc. abound\cite{dewitt}. \ 

The vacuum effects which we computed scale with Planck's constant and so are
very small. \ In order to have a practical spacecraft based on quantum
vacuum properties, it would probably have to be based on phenomena that
scale as $\hbar^{0}.$ \ By itself this requirement does not guarantee a
large enough magnitude, but it helps\cite{reynaud}. \ \ New methods of
modifying the quantum vacuum boundary conditions may be needed to generate
the large changes in energy or momentum required if "vacuum engineering" as
proposed in this paper is ever to be practical. \ For example, the vacuum
energy density between parallel plates is simply not large enough for our
engineering purposes. \ Energy densities that are orders of magnitude
greater are required. \ Such high energy density regions may be possible, at
least in some cases. \ For example, a region appeared in the 1 dimensional
dynamic system in which the energy density was below that of the Casimir
parallel plate region\cite{law}. \ More effective ways of transferring
momentum to the quantum vacuum than the adiabatic Casimir effect are
probably necessary if a spacecraft is to be propelled using the vacuum.

A proposal was made recently to measure the inertial mass shift in a
multilayer Casimir cavity, which consists of $10^{6}$ layers of metal $100$
nm thick, $35$ $\unit{cm}$ in diameter, alternating with films of silicon
dioxide $5$ nm thick\cite{calonni}. \ The mass shift is anticipated to arise
from the decrease in the vacuum energy between the parallel plates. \ \ A
calculation shows that the mass shift for the proposed cavity is at or just
beyond the current limit of detectability. \ It appears that if quantum
vacuum engineering of spacecraft is to become practical, and the dreams of
science fiction writers are to be realized, we may need to develop new
methods to be able to manipulate changes in vacuum energy densities that are
near to the same order of magnitude as mass energy densities. \ Then we
would anticipate being able to shift inertial masses by a significant
amount. \ \ Since mass shifts in computations are often formally infinite,
perhaps such developments are not forbidden. \ With large mass shifts one
might be able to build a structure that had a small or zero inertial mass,
which could be readily accelerated. \ Further, one could alter the curvature
of space-time is mesoscopic ways.

\begin{acknowledgments}
We gratefully acknowledge helpful comments and suggestions from Dan Cole,
and thank Gabriel Barton, Carlos Villarreal Lujan, Claudia Eberlein, and
Paulo Neto for answering questions about their work relavent to this paper.
\ We would like to express our appreciation to the NASA Breakthrough
Propulsion Physics Program and Marc Millis for the support of this work. \
We are sad to report that Robert L\ Forward died during the process of
preparing this paper.
\end{acknowledgments}

\newpage


\begin{thebibliography}{99}
\bibitem{bpp} Breakthrought Propulsion Physics Program of NASA,
http://www.grc.nasa.gov/WWW/bpp/.

\bibitem{clarke} Arthur C. Clarke, personal communication. \ See the
acknowledgements in "The Songs of Distant Earth."\ \ Numerous science
fiction writers, including Clarke, Asimov, and Sheffield have based
spacecraft on the quantum vacuum.

\bibitem{thorne} Morris M S and Thorne K S, "Wormholes in spacetime and
their use for interstellar travel: a tool for teaching general relativity,"
Am. J. Phys. 56, 395-412 (1988).

\bibitem{lamb} W. Lamb and R. Retherford, \textquotedblleft Fine structure
of the hydrogen atom by a microwave method,\textquotedblright\ Phys. Rev. 
\textbf{72}, 241 (1947).

\bibitem{bethe} H. Bethe, \textquotedblleft The Electromagnetic Shift of
Energy Levels,\textquotedblright\ Phys. Rev. \textbf{72}, 339 (1948).

\bibitem{cas1} H. B. G. Casimir, "On the attraction between two perfectly
conducting plates"Proc. K. Ned. Akad. Wet. \textbf{51} 793-5 (1948).

\bibitem{lamoroux} S. Lamoreaux, \textquotedblleft\ Measurement of the
Casimir force between conducting plates,\textquotedblright\ Phys. Rev. Lett. 
\textbf{78}, 5 (1997).

\bibitem{mohideen} U. Mohideen and A. Roy,\textquotedblleft Precision
Measurement of the Casimir Force from 0.1 to 0.9 micron\textquotedblright,
Phys. Rev. Lett. \textbf{81}, 4549 (1998).

\bibitem{klim} G. L. Klimchitskaya, A. Roy, U. Mohideen, and V. M.
Mostepanenko, "Complete roughness and conductivity corrections for Casimir
force measurement." Phys. Rev A\textbf{60} 3487 (1999).

\bibitem{mohideenpc} U. Mohideen, personal communication, 11/2002.

\bibitem{serry} F. Serry , D. Walliser, and J. Maclay, "The anharmonic
Casimir oscillator," J. Microelectromechanical Syst. \textbf{4,} 193 (1995).

\bibitem{serry2} M. Serry, D. Walliser, J. Maclay,"The role of the casimir
effect in the static deflection and stiction of membrane strips in
microelectromechanical systems (MEMS)," J. App. Phys. \textbf{84}, 2501
(1998).

\bibitem{buks} E. Buks and M. L. Roukes, "Stiction, adhesion, and the
Casimir effect in micromechanical systems," Phys. Rev. B \textbf{63}, 033402
(2001)

\bibitem{chan} H. Chan, V. Aksyuk, R. Kleiman, D. Bishop, and F. Capasso,
"Quantum Mechanical Actuation of Microelectromechanical Systems by the
Casimir Force," Science \textbf{291}, 1941 (2001)

\bibitem{chan2} H. B. Chan, V. A. Aksyuk, R. N. Kleiman, D. J. Bishop, and
F. Capasso, "Nonlinear Micromechanical Casimir Oscillator," Phys. Rev. Lett. 
\textbf{87}, 211801 (2001)

\bibitem{pwm} P. Milonni, \textit{The Quantum Vacuum. An Introduction to
Quantum Electrodynamics} (San Diego: Academic) 1994.

\bibitem{plunienreview} G. Plunien, B. M\"{u}ller, W. Greiner, "The Casimir
Effect," Physics Reports \textbf{134}, 87 (1986).

\bibitem{bordag} M. Bordag, U. Mohideen, V. Mostepanenko, "New Developments
in the Casimir Effect," Physics Reports \textbf{353}, 1 (2001).

\bibitem{cas3} Casimir H B G, Some remarks on the history of the so-called
Casimir effect \textit{The Casimir Effect 50 Years Later} ed M Bordag
(Singapore: World Scientific, 1999), pp 3-9.

\bibitem{schw} J. Schwinger J, L., DeRaad Jr L , and K. A. Milton, "Casimir
effect in dielectrics," Ann Phys (NY) \textbf{115} 1-23 (1978).

\bibitem{dodonov} V. Dodonov and A. Klimov, "Generation and detection of
photons in a cavity with a resonantly oscillating boundary," Phys.Rev. A 
\textbf{53}, 2664 (1996).

\bibitem{Power} E. A. Power E A and T. Thirunamachandran, "Zero-point energy
differences and many-body dispersion forces," Phys. Rev. A\textbf{50,}
3929-39 (1994).

\bibitem{moore} G. Moore, "Quantum theory of the electromagnetic field in a
variable-length one-dimensional cavity,"J. Math. Phys. \textbf{11}, 2679
(1970). \ \ 

\bibitem{netoandmachado} P. A. Maia Neto and L. A. S. Machado, "Quantum
radiation generated by a moving miror in free space," Phys. Rev. A \textbf{54%
}, 3420 (1996).

\bibitem{birrellanddavies} N. Birrell and P. Davies, \textit{Quantum fields
in curved space}, (Cambridge University Press, Cambridge, England, 1984), p.
48; p. 102.

\bibitem{jaekelandreynaud} M. Jaekel and S. Reynaud, Quantum Opt. \textbf{4}%
, 39 (1992); J. Phys. (France) I \textbf{2}, 149 (1992). \ \ 

\bibitem{fordandvilenkin} L. Ford and A. Vilenkin, "Quantum radiation by
moving mirrors," Phy. Rev. D, \textbf{25}, 2569 (1982).

\bibitem{bartonandeberlein} G. Barton and C. Eberlein, "On quantum radiation
from a moving body with finite refractive index," Ann. Phys. \textbf{227},
222 (1993).

\bibitem{neto} P. Neto, "Vacuum radiation pressure on moving mirrors," J.
Phys. A: Math. Gen. \textbf{27}, 2167 (1994).

\bibitem{jaekel} M. Jaekel and S. \ Reynaud, "Quantum fluctuations of mass
for a mirror in vacuum," Phys. Lett. A \textbf{180}, 9 (1993).

\bibitem{bartonandcal} G. Barton and A. Calogeracos, "On the quantum
dynamics of a dispersive miror. I. Radiative shifts, radiation, and
radiative reaction,"Ann. Phys. \textbf{238}, 227 (1995).

\bibitem{unruh} W. Unruh, "Notes on black-hole evaporation," Phys. Rev. D 
\textbf{14}, 870 (1976)

\bibitem{davies} P. Davies, "Scalar particle particle production in
Schwartzchild and Rindler metrics,"J. Phys. A \textbf{8}, 609 (1975);

\bibitem{hacyan} S. Hacyan, A. Sarmiento, G. Cocho, F. Soto, \ "Zero-point
field in curved spaces," \ Phys. Rev. D \textbf{32}, 914 (1985).

\bibitem{claudiapc} Personal communication from Claudia Eberlein. \ Prof.
Eberlein also points out the need to consider the compressibility of the
mirror for the frequencies likely to be excited.

\bibitem{eberleinson} C. Eberlein, "Theory of quantum radiation observed as
sonoluminescence," Phys.Rev. A \textbf{53}, 2772 (1996).

\bibitem{reynaud} A. Lambrecht, M. Jaekel, and S. Reynaud, "Motion induces
radiation from a vibrating cavity,"Phys. Rev. Lett. \textbf{77}, 615 (1996).

\bibitem{plunien} G. Plunien, R. Sch\"{u}tzhold, and G. Soff, "Dynamical
Casimir effect at finite temperature," Phys. Rev. Let. \textbf{84}, 1882
(2000).

\bibitem{forward} R. L. Forward, Extracting electrical energy from the
vacuum by cohesion of charged foliated conductors," \ Phys. Rev. B\textbf{30}%
, 1700 (1984).

\bibitem{coleandputhoff} D. C. Cole and H. E. Puthoff, "Extracting energy
and heat from the vacuum," Phys. Rev. E \textbf{48}, 1562 (1993).

\bibitem{rueda} A. Rueda, Space Science \textbf{53}, 223 (1990).

\bibitem{colethermod} D. C. Cole, "Possible thermodynamic violations and
astrophysical issues for secular acceleration of electrodynamic particles in
the vacuum," Phys. Rev. E, \textbf{51 }, 1663 (1995).

\bibitem{boyer} T. H. Boyer, "Thermal effects of acceleration through random
classical radiation,"Phys. Rev. D \textbf{21}, 2137 (1980).

\bibitem{pinto} F. Pinto, "Engine cycle of an optically controlled vacuum
energy transducer," Phys. Rev. B \textbf{60}, 14740 (1999).

\bibitem{law} C. K. Law, "Resonance response of the vacuum to an oscillating
boundary,"Phys. Rev. Lett. \textbf{73}, 1931 (1994).

\bibitem{carlos} C. Villarreal, S. Hacyan, R. Jauregui, \ "Creation of
particles and squeezed states between moving conductors," Phys. Rev. A 
\textbf{52}, 594 (1995).

\bibitem{boston} M. Boston, "Simplistic propulsion analysis of a
breakthrough space drive for Voyager," Proc. STAIF-00(Space Technology and
Applications International Forum, January 2000, Albuquerque, NM, ed. M.
El-Genk American Institute of Physics, New York, 2000).

\bibitem{eberleinden} C. Eberlein, "Quantum radiation from density
variations in dielectrics," J. Phys. A: Math. Gen. \textbf{32}, 2583 (1999).

\bibitem{yablonovitch} E. Yablonovitch, "Accelerating reference frame for
electromagnetic waves in a rapidly growing plasma:
Unruh-Davies-Fulling-DeWitt radiation and the nonadiabatic casimir effect,"
Phys. Rev. Lett. \textbf{62}, 1742 (1989).

\bibitem{lozovik} Y. Lozovovik, V. Tsvetus, E. Vinogradovc, JEPT\ Lett. 
\textbf{61}, 723 (1995).

\bibitem{Ford} L. H. Ford and N. F. Svaiter,\ "Focusing vacuum
fluctuations," Phys. Rev. A\ \textbf{62}, 062105 (2000).

\bibitem{bezerra} V. Bezerra, G. Klimchitskaya, C. Romero, "Surface
impedance and the Casimir force," Phys. Rev. A \textbf{65}, 012111 (2001).

\bibitem{genet} C. Genet, A. Lambrecht, S. Reynaud, "Temperature dependence
of the Csimir force between metallic mirrors," Phys. Rev. A 62, 012110
(2000).

\bibitem{bartonspheresetc} G. Barton, "Perturbative Casimir energies of
dispersive spheres, cubes, and cylinders," J. Phys. A.: Math. Gen. \textbf{34%
}, 4083 (2001).

\bibitem{dewitt} B. DeWitt in \textit{Physics in the Making, }pp 247-272,
ed. A. Sarlemijn and M. Sparnaay (Elsevier, Netherlands) 1989.

\bibitem{calonni} E. Calonni, L.DiFiore, G. Esposito, L. Milano, L. Rosa,
"Vacuum fluctuation force on a rigid Casimir cavity in a gravitational
field," in press Physics Letters A .
\end{thebibliography}
\end{document}